\documentclass[letterpaper, 10 pt, conference]{ieeeconf}
\IEEEoverridecommandlockouts                              
\usepackage[utf8]{inputenc}
\usepackage[T1]{fontenc}
\usepackage{graphicx, xcolor}
\usepackage{mathtools}
\usepackage{hyperref}
\usepackage{times}
\usepackage{fancyhdr,graphicx,amsmath,amssymb}
\usepackage[ruled,vlined]{algorithm2e}
\include{pythonlisting}

\title{\LARGE \bf
Compact Merkle Multiproofs
}

\author{Lum Ramabaja \\ \href{mailto:lum@bloomlab.io}{lum@bloomlab.io} 
   \and Arber Avdullahu \\ \href{mailto:lum@bloomlab.io}{arber@bloomlab.io} 
}

\begin{document}

\maketitle

\begin{abstract}

The compact Merkle multiproof is a new and significantly more memory-efficient way to generate and verify sparse Merkle multiproofs. A standard sparse Merkle multiproof requires to store an index for every non-leaf hash in the multiproof. The compact Merkle multiproof on the other hand requires only $k$ leaf indices, where $k$ is the number of elements used for creating a multiproof. This significantly reduces the size of multiproofs, especially for larger Merkle trees.

\end{abstract}

\section{Related Work}
We received a lot of useful feedback from the 
\href{https://news.ycombinator.com/item?id=22364786}{\textcolor{blue}{Hacker News}} community, after we put the first version of this paper was put online. Even though we were not able to find similar work to the compact Merkle multiproof in the literature, it turns out two people already worked on almost identical proposals before. Luke Champine contributed to a repository more than a year ago with almost the same concept as the one we will propose in this paper \cite{AddGitLab}. The only difference between our approach and Champine's approach, is that his implementation uses a right-to-left technique for building a multiproof, whereas our approach uses a bottom-up technique to construct the multiproof. Besides Champine's work, it is also worth noting Pieter Wuille's contributions to the bitcoin protocol \cite{AddGitHub}. Wuille's multiproof approach is different from both our work and Champine's work, as it does not rely on leaf indices for multiproof construction. The end result however is just as compact, and because of that deserves mentioning.

\section{Introduction}
\label{s:intro}
In this paper we will introduce the compact Merkle multiproof, a more efficient way to compute and transmit sparse Merkle multiproofs \cite{UnderstandingTechnology}. To understand how the compact Merkle multiproof works, we have to understand first how Merkle trees function, and what sparse Merkle multiproofs are. This is why in this brief introduction, we are going to briefly explain both concepts, before continuing with the compact Merkle multiproof algorithm.

\subsection{Merkle trees}

\begin{figure}[h]
\centering\includegraphics[width=0.7\linewidth]{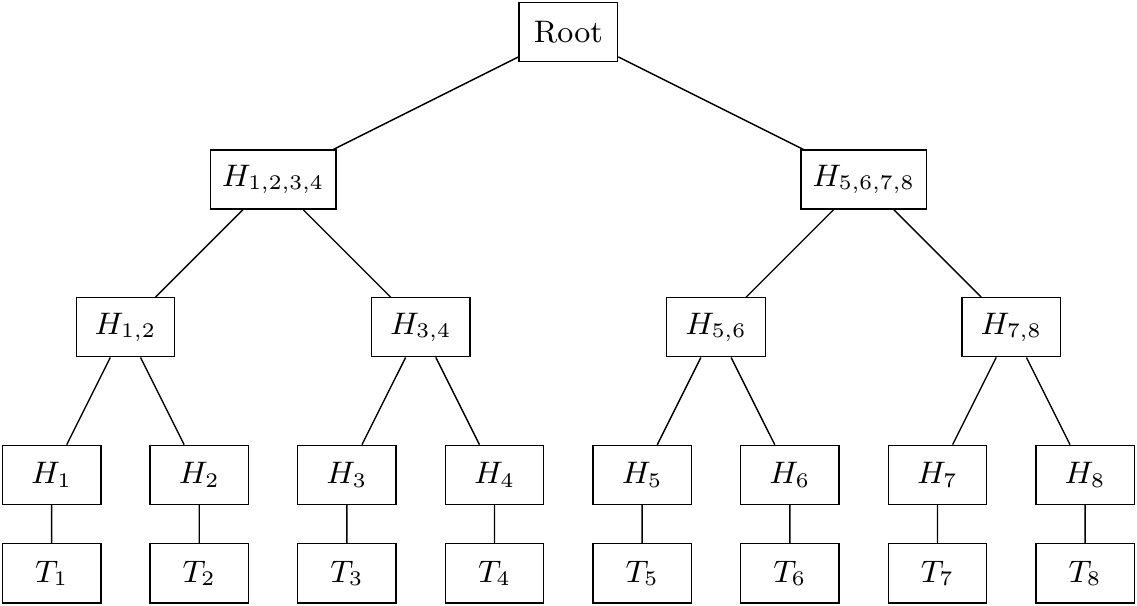}
\caption{Depiction of a Merkle tree. The leaf nodes, i.e. the elements of a Merkle tree, are written as '$T_i$. The non-leaf nodes are written as $H_j$.}
\label{fig:MT}
\end{figure}

A Merkle tree is a binary tree in which all leaf nodes (i.e. the Merkle tree's elements) are associated with a cryptographic hash, and all none-leaf nodes are associated with a cryptographic hash, that is formed from the hashes of its child nodes (as shown in figure \ref{fig:MT}).  

The Merkle tree is a data structure that allows for bandwidth-efficient and secure verification of elements in a list. It is used to verify the presence of elements in and between computers, without having to send the whole list of elements to another computer. Merkle trees have found a variety of uses cases: They are used in peer-to-peer systems to verify if the integrity of data blocks \cite{BenetIPFS-Content3}, for batch signing of time synchronisation requests \cite{RoughtimeGoogle}, for transaction verification in blockchain systems \cite{NakamotoBitcoin:System}, and more. 

\begin{figure}[h]
\centering\includegraphics[width=0.7\linewidth]{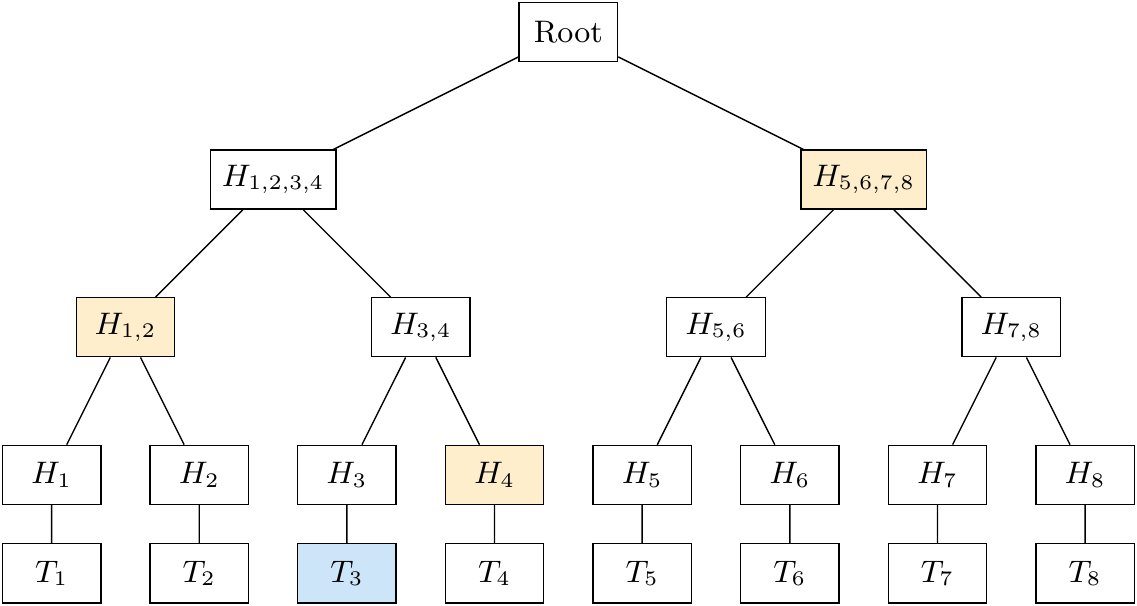}
\caption{Depiction of a Merkle tree, with the Merkle proof (shown in orange) for a given element ($T_3$).}
\label{fig:MT1}
\end{figure}

To verify that an element is present in the Merkle tree, a series of hashes are provided. the series of hashes is also known as a Merkle proof. By sequentially hashing an element hash with the provided Merkle proof, one can recreate the the Merkle root of the Merkle tree (as shown in figure \ref{fig:MT1}). If an element is present in a list and the Merkle proof is correct, then the end result of the sequential hashing will be the Merkle root. The recipient of the Merkle proof thus already has to have a copy of the Merkle root, before verifying the integrity of a Merkle proof. As an example, by periodically storing Merkle roots, a verifier will be able to prove that some data is still unaltered. 

\subsection{Sparse Merkle Multiproofs}
\begin{figure}[h]
\centering\includegraphics[width=0.7\linewidth]{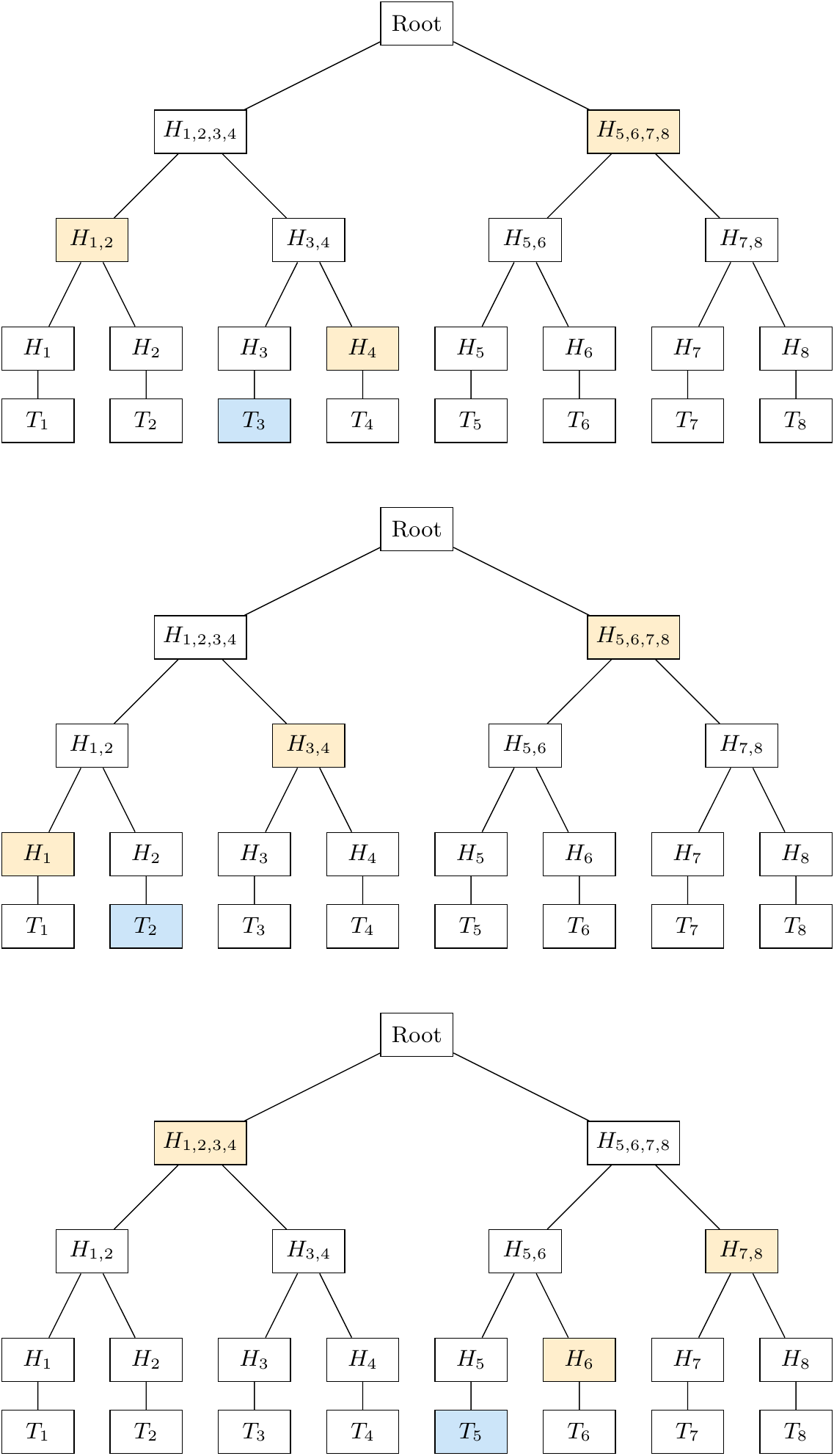}
\caption{Three Merkle proofs for three different elements.}
\label{fig:Merkle_three}
\end{figure}

A sparse Merkle multiproof (not to be confused with sparse Merkle trees) is a more efficient Merkle proof, for when it is necessary to prove the presence of a multiple elements that are in the same Merkle tree \cite{UnderstandingTechnology}. Let's take figure \ref{fig:Merkle_three} as an example to better understand what this means. To prove that three different elements are present in a Merkle tree, we could compute three separate Merkle proofs and verify the presence of each element separately. In the provided example, a node would need nine hashes in total to verify the presence of three elements. 

\begin{figure}[h]
\centering\includegraphics[width=0.7\linewidth]{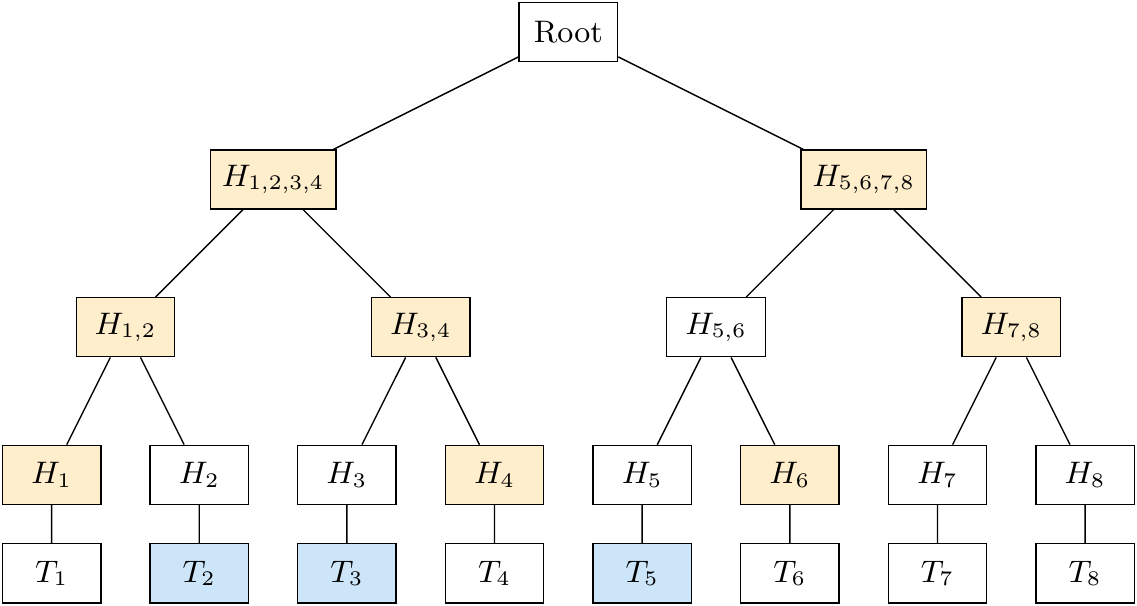}
\caption{Three overlapped Merkle proofs.}
\label{fig:Merkle_overlapped}
\end{figure}

\begin{figure}[h]
\centering\includegraphics[width=0.7\linewidth]{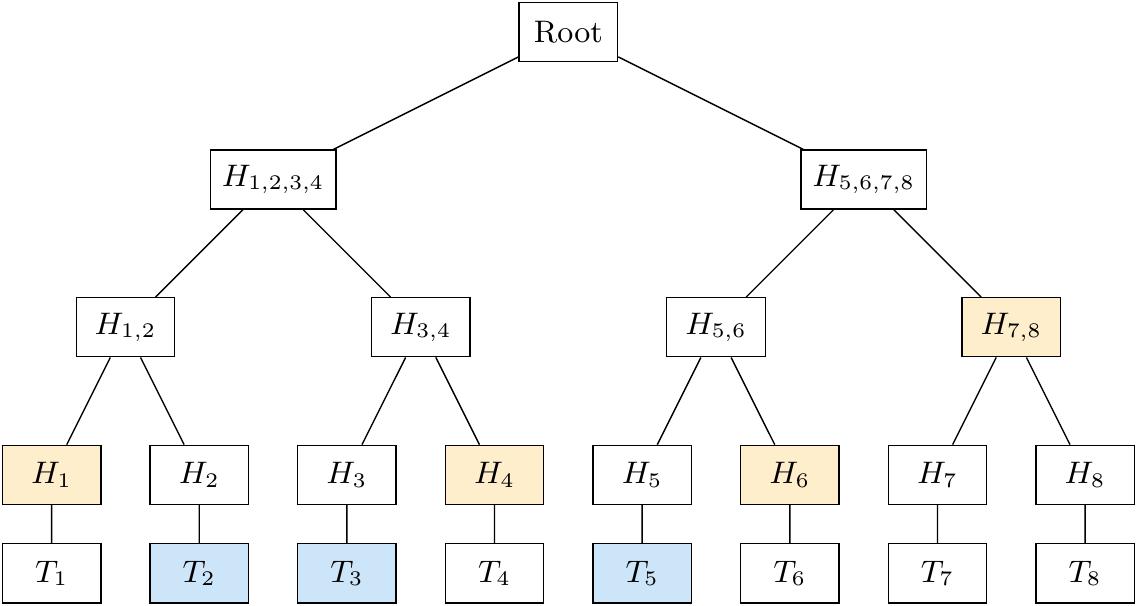}
\caption{An illustration of a Merkle multiproof.}
\label{fig:Merkle_multiproof}
\end{figure}

By using a sparse Merkle multiproof however, we can drop the number of hashes significantly. When overlapping the three the Merkle proofs from figure \ref{fig:Merkle_three} (as shown in figure \ref{fig:Merkle_overlapped}), we can see that many of the hashes can in fact be recreated by previous hashes. Instead of using three separate Merkle proofs that consist of nine hashes in total, one can prove the presence of the three elements with only four hashes (as shown in figure \ref{fig:Merkle_multiproof}. This simple trick is also known as a sparse Merkle multiproof. 

\begin{figure}[h]
\centering\includegraphics[width=0.7\linewidth]{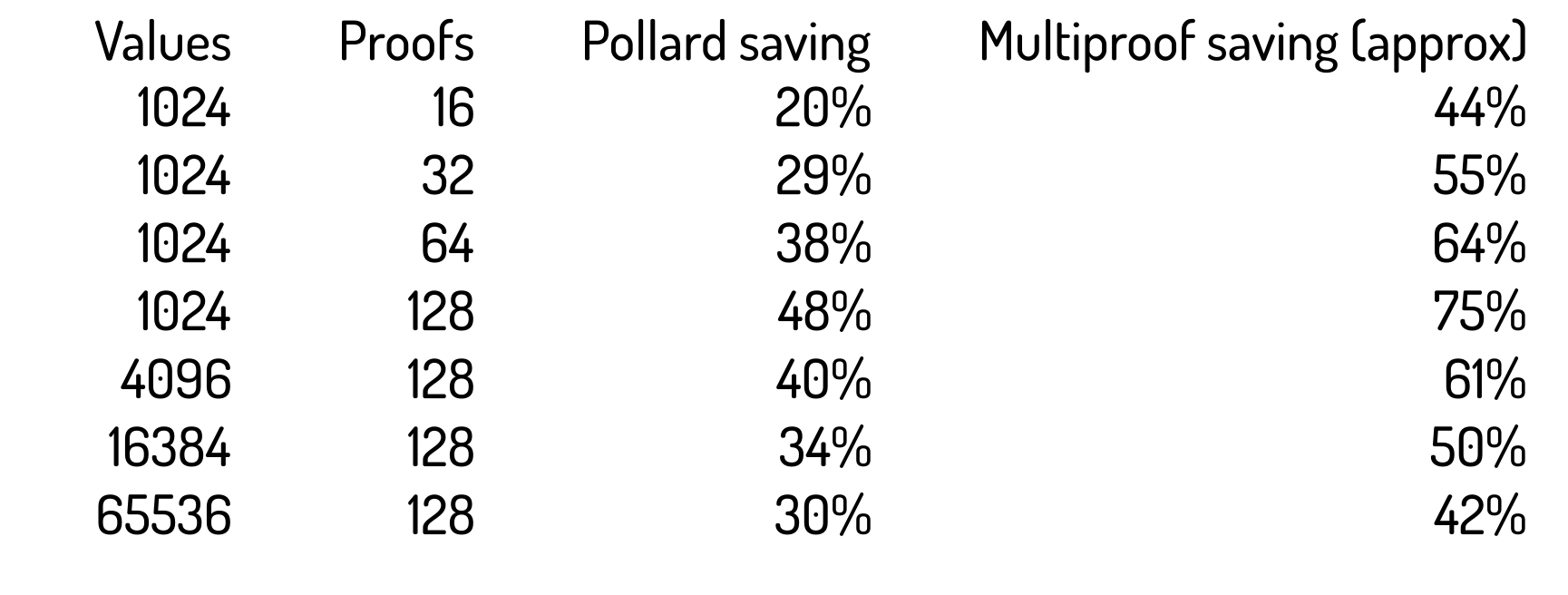}
\caption{Table taken from Jim McDonald's wonderful article "Understanding sparse Merkle multiproofs" \cite{UnderstandingTechnology}. Space saving for Merkle pollards and spare Merkle multiproofs over simple Merkle proofs.}
\label{fig:McDonalds_picture}
\end{figure}

Using sparse Merkle multiproofs over standard Merkle proofs can have enormous space savings in certain scenarios, as shown in figure \ref{fig:McDonalds_picture}. There is however one important problem with current sparse Merkle multiproof implementations that we thought needs addressing, today's implementations require additional data besides the multiproof \cite{UnderstandingTechnology}. Today's sparse Merkle multiproofs require to store the hash indices for every non-leaf node. In other words, for every hash in a multiproof, we need an index to figure out the order of computations in order to reconstruct a given Merkle root. One could argue that the necessity for additional data defeats the purpose of using a sparse Merkle multiproof, or at least significantly limits its potential. This precise issue is what the compact Merkle multiproof solves.

\section{The Compact Merkle Multiproof}
The compact Merkle multiproof is a special technique to generate and verify sparse Merkle multiproofs, without the need for non-leaf index information. A standard sparse Merkle multiproof requires to store an index for every non-leaf hash in the multiproof, the compact Merkle multiproof on the other hand requires only $k$ leaf indices (or in the case of the Bloom tree \cite{Ramabaja2020TheTree}, $k$ Bloom filter chunks), where $k$ is the number of elements used for creating a multiproof. This significantly reduces the size of multiproofs, especially for larger Merkle trees. 

\begin{figure*}[tp]
\centering\includegraphics[width=0.8\linewidth]{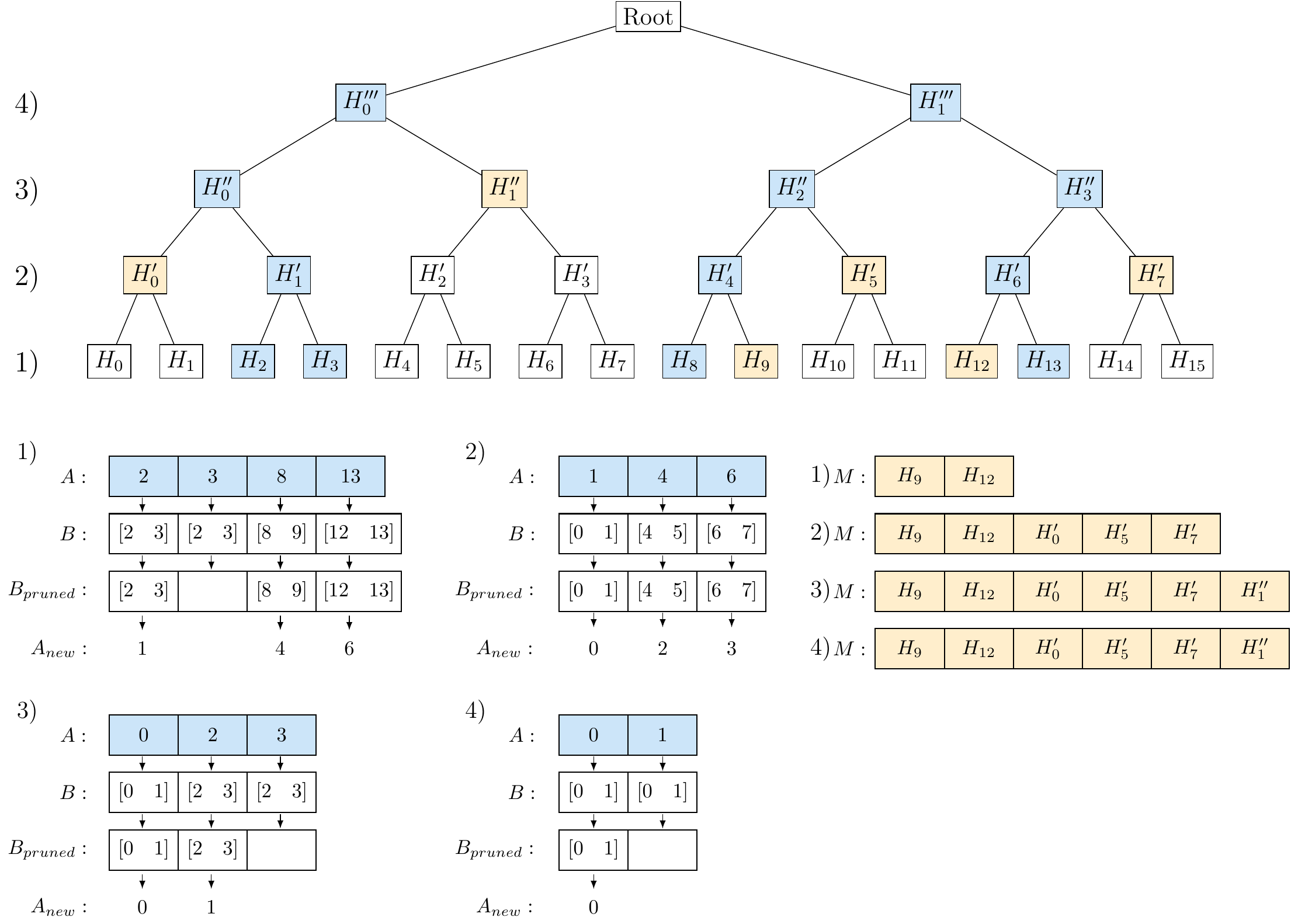}
\caption{An illustration of the compact Merkle multiproof generation procedure. Orange boxes represent the hashes of the Merkle proof. Blue boxes represent the indices of $A$ in every Merkle layer. In each iteration, we append hashes to $M$, until the tree root is reached. For a more detailed description of the procedure, refer to subsection \ref{sub:MerkleMultiproofGeneration}. }
\label{fig:multiproof_generation}
\end{figure*}

\begin{figure*}[bp]
    \centering
    \includegraphics[width=0.8\linewidth]{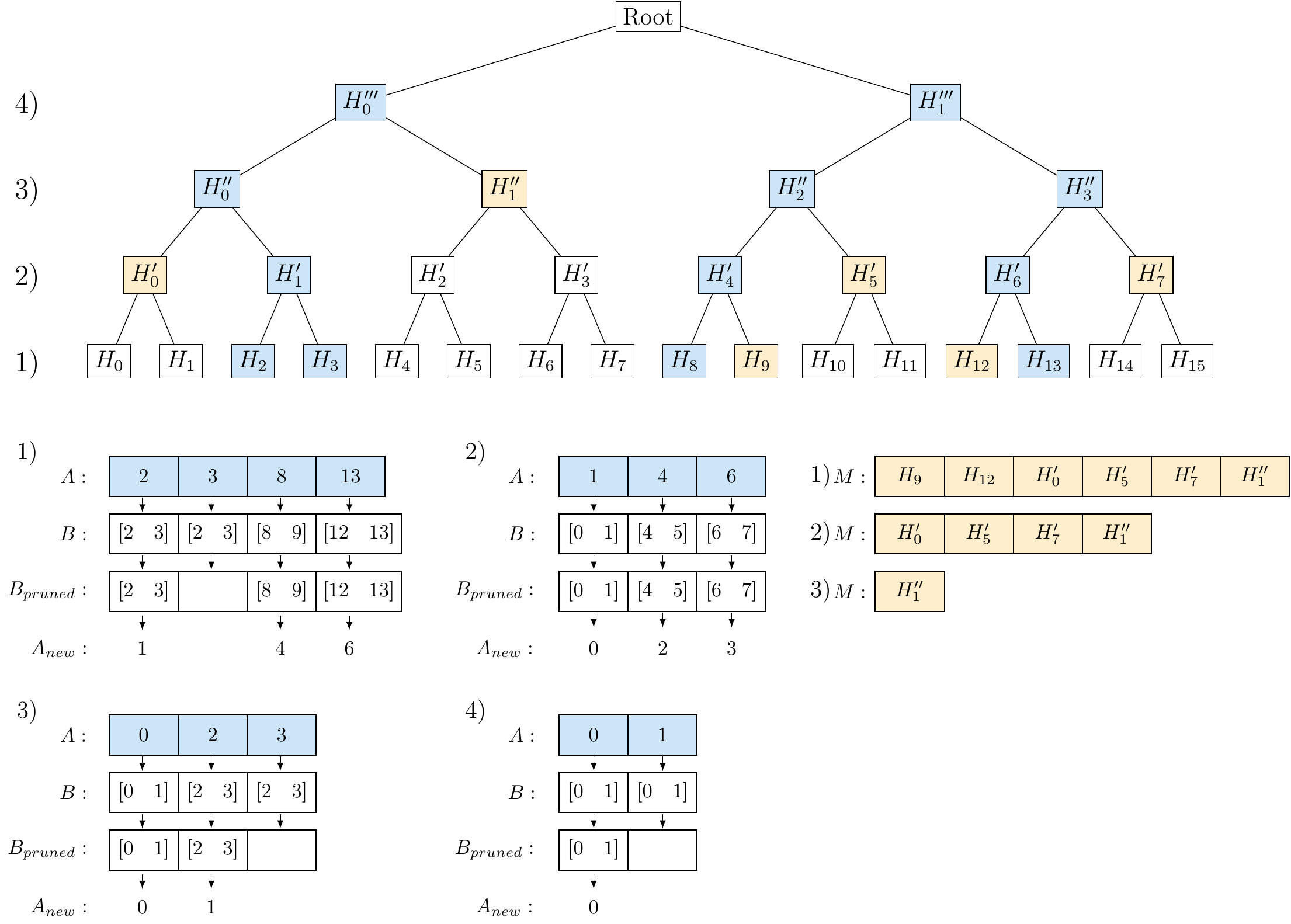}
    \caption{An illustration of the compact Merkle multiproof verification procedure. Orange boxes represent the hashes of the Merkle proof. Blue boxes represent the indices of $A$ in every Merkle layer. In each iteration, we hash elements between the element hashes in $E$ and $M$ until no element is left in $M$. For a more detailed description of the procedure, refer to subsection \ref{sub:MerkleMultiproofVerification}. }
    \label{fig:multiproof_verification}
\end{figure*}

In the next subsections we will explain how to generate and verify a compact Merkle multiproof for a Merkle tree. It is important to note that the compact Merkle multiproof technique works with other kinds of Merkle trees as well, such as Bloom trees, sparse Merkle trees, sorted Merkle trees, etc. We are going to take figure \ref{fig:multiproof_generation} and \ref{fig:multiproof_verification} as references to better understand how compact Merkle multiproofs are generated and verified.

\subsection{Compact Merkle Multiproof for Merkle Trees}

\subsubsection{Compact Merkle Multiproof Generation}
\label{sub:MerkleMultiproofGeneration}
Every leaf node has an index from 0 to $N$, where $N$ is the total number of leaves in a Merkle tree. We first determine the index for every element that takes part in the multiproof. In the case of our example in figure \ref{fig:multiproof_generation}, that would be indices $[2, 3, 8, 13]$. Let's name this array $A$ and let's name the "Merkle layer" on which we operate as $L$ (The Merkle Layer at the beginning are simply the leaf nodes of the tree). After determining these indices, we run the following steps recursively until termination:
\begin{enumerate}
    \item[--] For each of the indices in $A$, take the index of its immediate neighbor in layer $L$, and store the given element index and the neighboring index as a pair of indices (an "immediate neighbor" is the leaf index right next to a target leaf index that shares the same parent). In the first iteration of our example in figure \ref{fig:multiproof_generation}, we end up with an array of the form $[[2,3],[2,3],[8,9],[12,13]]$. Let's name this array $B$
    \item[--] Remove any duplicate from $B$. In the first iteration of our example in figure \ref{fig:multiproof_generation}, $B$ would end up to be of the form $[[2,3],[8,9],[12,13]]$. In figure \ref{sub:MerkleMultiproofGeneration} we refer to this as $B_{pruned}$.
    \item[--] Take the difference between the set of indices in $B_{pruned}$ and $A$ and append the hash values for the given indices, for the given Merkle layer to the multirpoof $M$. In the first iteration of our example in figure \ref{fig:multiproof_generation}, we would end up with the indices $[9, 12]$, which are the indices for $H_{9}$ and $H_{12}$. Append $H_{9}$ and $H_{12}$ to the multiproof $M$.
    \item[--] We take all the even numbers from $B_{pruned}$, and divide them by two. We assign the newly computed numbers to $A$. In the first iteration of our example in figure \ref{fig:multiproof_generation}, $A$ would end up to be of the form $[1, 4, 6]$.
    \item[--] Go one layer up the tree. Assign that layer to $L$. Each layer in the tree is indexed from 0 to $N$, where $N$ is the size of that layer.
    \item[--] Repeat the above steps with the newly assigned variables $A$ and $L$ until you reach the root of the tree. 
\end{enumerate}{}
The proof at the end must contain the indices of the elements used for the multiproof, as well as the gathered hashes inside $M$.

\subsubsection{Compact Merkle Multiproof Verification}
\label{sub:MerkleMultiproofVerification}
For a compact Merkle multiproof verification, we require the indices of the elements used for a multiproof (Let's name this array $A$), the corresponding hashes for the elements used for a multiproof, as well as the hashes of the multiproof $M$. Array $A$ in case of our example \ref{fig:multiproof_verification} would be $[2,3,8,13]$. We first need to sort the $k$ element hashes in increasing order according to $A$. Let's name this sorted array $E$. To verify a generated multiproof, we run the following steps recursively until termination:
\begin{enumerate}
    \item[--] For each of the indices in $A$, take the index of its immediate neighbor, and store the given element index and the neighboring index as a pair of indices (an "immediate neighbor" is the leaf right next to a target leaf that shares the same parent). In the first iteration of our example in figure \ref{fig:multiproof_verification}, we would end up with an array of the form $[[2,3],[2,3],[8,9],[12,13]]$. Let's name this array $B$.
    \item[--] $B$ will always have the same size as $E$. After computing $B$, we check for duplicate index pairs inside it. If two pairs are identical, we hash the corresponding values (that have the same indices) inside $E$ with one another. If an index pair has no duplicates, we hash the corresponding value inside $E$ with the first hash inside $M$. If a value inside $M$ was used, we remove it from $M$. All the newly generated hashes are assigned to a new $E$ that will be used for the next iteration.
    \item[--] We take all the even numbers from $B_{pruned}$, and divide them by two. We assign the newly computed numbers to $A$. In the first iteration of our example in figure \ref{fig:multiproof_verification}, $A$ would end up to be of the form $[1, 4, 6]$
    \item[--] Repeat the above steps until $M$ has no elements anymore. 
\end{enumerate}{}
At the end of this procedure, $E$ will have a single value, the Merkle root of the tree. If the final value is not equal to the stored Merkle root, the verifier knows that the proof is invalid.

\section{Conclusion}
We showed a new way how to compute more memory-efficient Merkle multiproofs. The compact Merkle multiproof can generate and verify sparse Merkle multiproofs, without the need for non-leaf index information. A standard sparse Merkle multiproof requires to store an index for every non-leaf hash in the multiproof, the compact Merkle multiproof on the other hand requires only $k$ leaf indices, where $k$ is the number of elements used for creating a multiproof. This significantly reduces the size of multiproofs, especially for larger Merkle trees. The compact Merkle multiproof technique can be applied to a various Merkle tree variants, such as the Bloom tree, sparse Merkle tree, etc.

\section{Future Work}
We have an implementation of the compact Merkle multiproof for our Bloom tree package (which can be found on the Bloom Lab's \href{https://github.com/labbloom/bloom-tree}{\textcolor{blue}{github}} page).
In future work, we are going to show how one can combine Bloom trees that use compact Merkle multiproofs, with distributed Bloom filters \cite{Ramabaja2019TheFilter} to create an ''interactive Boom proof''. We will show how the interactive Bloom proof can be used to build a new kind of blockchain architecture, that requires one magnitude less storage, while still allowing nodes to independently verify transaction validity. The efficiency of the compact Merkle multiproof procedure will play an integral part in this setup.


\addtolength{\textheight}{-12cm}   



\bibliographystyle{IEEEtran}

\begin{thebibliography}{1}
\providecommand{\url}[1]{#1}
\csname url@samestyle\endcsname
\providecommand{\newblock}{\relax}
\providecommand{\bibinfo}[2]{#2}
\providecommand{\BIBentrySTDinterwordspacing}{\spaceskip=0pt\relax}
\providecommand{\BIBentryALTinterwordstretchfactor}{4}
\providecommand{\BIBentryALTinterwordspacing}{\spaceskip=\fontdimen2\font plus
\BIBentryALTinterwordstretchfactor\fontdimen3\font minus
  \fontdimen4\font\relax}
\providecommand{\BIBforeignlanguage}[2]{{%
\expandafter\ifx\csname l@#1\endcsname\relax
\typeout{** WARNING: IEEEtran.bst: No hyphenation pattern has been}%
\typeout{** loaded for the language `#1'. Using the pattern for}%
\typeout{** the default language instead.}%
\else
\language=\csname l@#1\endcsname
\fi
#2}}
\providecommand{\BIBdecl}{\relax}
\BIBdecl

\bibitem{AddGitLab}
\BIBentryALTinterwordspacing
``{add support for multi-range proofs (bae5b547) {\textperiodcentered} Commits
  {\textperiodcentered} NebulousLabs / merkletree {\textperiodcentered}
  GitLab}.'' [Online]. Available:
  \url{https://gitlab.com/NebulousLabs/merkletree/-/commit/bae5b547495f9dddbd4ddeecfdcb00cb89d99d76}
\BIBentrySTDinterwordspacing

\bibitem{AddGitHub}
\BIBentryALTinterwordspacing
``{Add CPartialMerkleTree {\textperiodcentered} bitcoin/bitcoin@4bedfa9
  {\textperiodcentered} GitHub}.'' [Online]. Available:
  \url{https://github.com/bitcoin/bitcoin/commit/4bedfa9223d38bbc322d19e28ca03417c216700b}
\BIBentrySTDinterwordspacing

\bibitem{UnderstandingTechnology}
\BIBentryALTinterwordspacing
``{Understanding sparse Merkle multiproofs | Weald Technology}.'' [Online].
  Available:
  \url{https://www.wealdtech.com/articles/understanding-sparse-merkle-multiproofs/}
\BIBentrySTDinterwordspacing

\bibitem{BenetIPFS-Content3}
J.~Benet, ``{IPFS-Content Addressed, Versioned, P2P File System (DRAFT 3)},''
  Tech. Rep.

\bibitem{RoughtimeGoogle}
\BIBentryALTinterwordspacing
``{roughtime - Git at Google}.'' [Online]. Available:
  \url{https://roughtime.googlesource.com/roughtime}
\BIBentrySTDinterwordspacing

\bibitem{NakamotoBitcoin:System}
\BIBentryALTinterwordspacing
S.~Nakamoto, ``{Bitcoin: A Peer-to-Peer Electronic Cash System},'' Tech. Rep.
  [Online]. Available: \url{www.bitcoin.org}
\BIBentrySTDinterwordspacing

\bibitem{Ramabaja2020TheTree}
\BIBentryALTinterwordspacing
L.~Ramabaja and A.~Avdullahu, ``{The Bloom Tree},'' 2 2020. [Online].
  Available: \url{http://arxiv.org/abs/2002.03057}
\BIBentrySTDinterwordspacing

\bibitem{Ramabaja2019TheFilter}
\BIBentryALTinterwordspacing
------, ``{The Distributed Bloom Filter},'' 10 2019. [Online]. Available:
  \url{http://arxiv.org/abs/1910.07782}
\BIBentrySTDinterwordspacing

\end{thebibliography}

\end{document}